# Plasmonic Mediated Atomically Engineered 2D Aluminium Quasicrystals for Dopamine Biosensing

*Saswata Goswami[†], Guilherme S. L. Fabris[†], Diganta Mondal, Raphael B. de Oliveira, Anyesha Chakraborty, Thakur Prasad Yadav, Nilay Krishna Mukhopadhyay, Samit K. Ray*, Douglas S. Galvão*, and Chandra Sekhar Tiwary**

S. Goswami, A. Chakraborty
School of Nano Science and Technology, Indian Institute of Technology, Kharagpur, West Bengal-721302, India

G. S. L. Fabris, D. S. Galvão
Applied Physics Department and Center for Computational Engineering & Sciences, State University of Campinas, Campinas, São Paulo 13083-970, Brazil
E-mail: galvao@ifi.unicamp.br

D. Mondal
Department of Industrial & Applied Chemistry, Ramakrishna Mission Vidyamandira (Affiliated to University of Calcutta), West Bengal-711202, India

R. B. Oliveira
Department of Materials Science and NanoEngineering, Rice University, Houston, TX, USA

T.P. Yadav
Department of Physics, Faculty of Science, University of Allahabad, Prayagraj 211002 UP, India

N. K. Mukhopadhyay
Department of Metallurgical Engineering, Indian Institute of Technology (BHU), Varanasi 221005, UP, India

S. K. Ray
Department of Physics, Indian Institute of Technology Kharagpur, West Bengal 721302, India
Email: physkr@phy.iitkgp.ac.in

C. S. Tiwary
Department of Metallurgical and Materials Engineering, Indian Institute of Technology Kharagpur, West Bengal 721302, India
E-mail: chandra.tiwary@metal.iitkgp.ac.in

[†] Authors with equal contribution.

## Abstract

Dopamine levels are linked to neurological illnesses like Parkinson's and Alzheimer's. Thus, reliable and sensitive detection of dopamine is crucial for early diagnosis and surveillance of neurodegenerative diseases. Non-noble-metal-based nanomaterials are ideal for light-mediated sensing of organic molecules. Among these 2D quasicrystal structures consisting of five elements 2D ($Al_{70}Co_{10}Fe_5Ni_{10}Cu_5$ (Al QC)) provides active sites due to their high surface-to-volume ratio, making them excellent for organic chemical sensing. Here, we propose a simple, label-free, spatial self-phase-modulation (SSPM)-based sensing method in liquid form. SSPM-based time evolution of the diffraction pattern for varied mixing levels of a 1100 ppb dopamine solution shows a shift in active solution 2D Al QC. The 1100 ppb solution shows a different $dN/dI$ value, indicating a change in the nonlinear refractive index. Time-evolution analysis is used to calculate sensitivities to changes in the nonlinear refractive index and time constant. The SPR-activated 2D Al QC nanostructure is used to demonstrate dopamine-sensing methods and to perform qualitative and quantitative evaluations. The SSPM-based sensing has been further compared using other optical-based sensing methods such as Raman Spectroscopy, UV-Vis Spectroscopy, and FTIR Spectroscopy. The experimental observations are also explained using DFT-based simulations. The current SSPM method can be used for rapid, large-scale medical diagnostics.

## 1. Introduction

Sensitive detection of biomolecules such as proteins, nucleic acids, neurotransmitters, and metabolites is critical for biomedical diagnostics, environmental monitoring, and food safety. Optical spectroscopy techniques have emerged as powerful analytical tools because they provide rapid, label-free, and non-destructive detection of biomolecular species. Among these techniques, Raman spectroscopy, Fourier Transform Infrared (FTIR) spectroscopy, and ultraviolet (UV-Vis) spectroscopy are widely used to probe molecular interactions and the structural properties of biomolecules. In biosensing applications, plasmonic materials are widely used to improve the performance of spectroscopic techniques such as Raman, FTIR, and UV-Vis spectroscopy. For example, surface-enhanced Raman scattering intensity enables the detection of biomolecules at very low concentrations.[1] Similarly, plasmonic nanostructures can enhance the infrared absorption (SEIRA) spectroscopy, allowing improved detection of molecular vibrational signatures. In UV-Vis spectroscopy, plasmonic nanoparticles exhibit characteristic absorption bands, and biomolecule binding can shift the plasmon resonance peak, providing the basis for colorimetric sensing.[2] Because of their ability to amplify optical signals and provide highly sensitive detection, plasmon-activated materials have become key components in modern biosensors for detecting biomolecules, including proteins, nucleic acids, and neurotransmitters.[3] Their integration with advanced nanostructures and optical platforms continues to drive the development of next-generation sensing technologies for biomedical diagnostics and environmental monitoring. Two-dimensional materials have recently been explored for molecular and biosensing applications due to several key factors, including a high surface-to-volume ratio, tunable electronic states, and surface active sites that facilitate analyte-2D material interactions. Among these, quasicrystals have been used for next-generation plasmonic and optoelectronic applications. Quasicrystals are known to be intermetallic phases exhibiting no translational symmetry and a long-range quasi-periodic nature. This non-periodic atomic arrangement yields dense k-space distribution in the Brillouin zone and unusual light-matter activity. Recent investigations show that multicomponent quasicrystals can be exfoliated into 2D counterparts, facilitating nanoscale light-matter interactions, catalytic applications, and surface-enhanced phenomena. Mandal *et al.* reported successful exfoliation of Ti-based QC ($Ti_{45}Zr_{38}Ni_{17}$) and its plasmonic response capability, and dopamine sensing capability in the nanomolar range.[4] Kumbhakar *et al.* investigated that 2D Al-based quasicrystal ($Al_{70}Co_{10}Fe_5Ni_{10}Cu_5$) can exhibit strong light localization, plasmonic scattering, and enhanced nonlinear optical responses, making it a suitable candidate for molecular sensing applications.[5] Recent literature studies show that multi-component quasicrystals with non-noble metals can outperform traditional 2D materials such as graphene and $MoS_2$, due to their unique properties under laser illumination and inherent plasmonic modes. Dopamine serves as a critical neurotransmitter and a biomarker for disorders such as Alzheimer's and Parkinson's disease. Structural similarity between dopamine and other molecules, such as uric acid and ascorbic acid, makes its detection challenging. Conventional dopamine sensors use enzymatic or electrochemical devices, which makes the sensing mechanism complex.

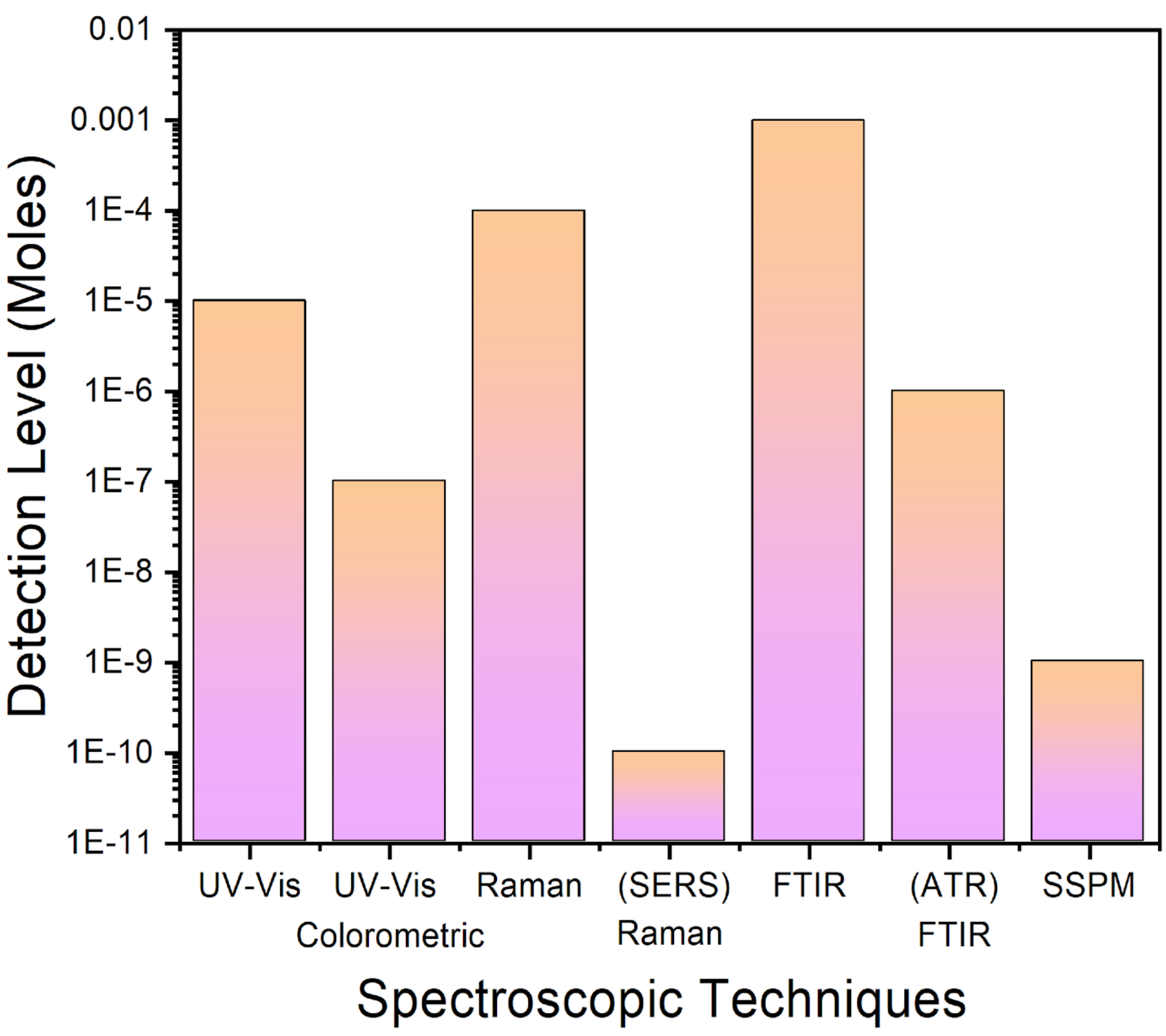

**Figure 1**. Comparative detection limit with spectroscopic techniques.

The process also requires surface functionalization and doping with other foreign metals, further increasing complexity and rendering the stability of the 2D materials irretrievable. 2D $Al_{70}Co_{10}Fe_5Ni_{10}Cu_5$ quasicrystals offer an alternative approach due to the high density of reactive metal atoms on their surfaces; these surface atoms react with the phenolic group on the dopamine molecule, enabling label-free dopamine sensing. A prior study with a Ti-based quasicrystal shows measurable shifts in UV-vis absorbance, Raman spectra, and diffraction-based optical responses, establishing a mechanistic basis for QC-mediated biomolecular sensing.[4] Previously, photonic isolators [6, 7], photonic switches [8, 9], and logic gates [10, 11], as well as nonlinear devices, have been realized using SSPM spectroscopy. Surface plasmon-assisted biomolecule detection has never been realized; this work presents a new avenue for organic-molecule sensing using a plasmonic 2D quasicrystal nanostructure. **Figure 1** shows the detection limit for various spectroscopic techniques. In this work, an Aluminium-based quasicrystal ($Al_{70}Co_{10}Fe_5Ni_{10}Cu_5$) was further exfoliated in a 2D form via liquid-phase exfoliation. Structural characterization, including AFM, SEM, and TEM, was used to investigate the 2D nature of the Al QC. The SPR property was investigated using an optical microscope under dark-field conditions with 532 nm, 15-mW laser illumination. This plasmonic nature was verified through sensing mechanisms using traditional methods such as UV-Vis, Raman, and FTIR Spectroscopy. The interaction between dopamine and the 2D Al QC was investigated using the density functional tight-binding method, including isopropanol solvation effects. A novel method was developed using SSPM spectroscopy, exploiting plasmonically induced optical nonlinearity to detect different concentrations of dopamine. SSPM-based spectroscopy represents a robust visual method for sensing dopamine molecules at nanomolar levels.

## 2. Synthesis and Characterization

Polycrystalline alloy was fabricated from high-purity Al (99.95%), Co (99.90%), Fe (99.99%), Ni (99.5%) (Alfa Aesar), and Cu (99.5%) (Sigma-Aldrich). The constituent elements were melted five times to achieve homogeneity. Before alloy melting, titanium buttons were used to remove residual oxygen from the chamber. After oxygen removal from the chamber, the bulk material was encapsulated in a quartz tube under an inert argon atmosphere and annealed at 1000°C for 48 hr. The quasicrystal material was found to be brittle and easily converted to powder using a mortar and pestle. Atomically thin 2D nanostructures were produced by sonicating the annealed bulk samples in IPA using a pressurized ultrasonic reactor. For solvent-assisted exfoliation, 10 mg of QC powder was dispersed in 50 mL of IPA. The active sonication period of 4 hr. was performed in the bulk QC containing IPA solution. The intense sonication pulses generated the 2D Al Quasicrystal in a thin sheet-like nanostructure from the surface of the bulk powder. **Figure 2(a)** displays the characteristic X-ray diffraction (XRD) profile of $Al_{70}Co_{10}Fe_{5}Ni_{10}Cu_{5}$ decagonal quasicrystals of the bulk and 2D nanostructure, respectively.[12] The diffraction pattern of decagonal quasicrystals follows a unique symmetry-based indexing scheme. All XRD reflections were indexed using the least path criterion with redundant basis vectors. The absence of certain Bragg peaks in the exfoliated samples indicates that only specific planes, namely, $(10\overline{11}00)$ and $(10\overline{22}02)$, are present in the 2D Al QC nanostructures, dispersed in IPA. The peaks corresponding to the decagonal phase are linked to quasi-lattice parameters, $a_R$= 3.98 Å and c = 8.19 Å, which are somewhat greater than those of the bulk materials ($a_R$ = 3.91 Å, c = 8.12 Å). Noticeable peak broadening appears in the exfoliated decagonal quasicrystal near 2θ = 43.3°, whereas the bulk sample exhibits sharp XRD reflections. This broadening is attributed to increased disorder in the decagonal plane, arising from atomic relaxation or structural perturbations introduced during exfoliation. In the 2D Al QC nanostructure XRD spectrum, fewer planes are prominent, while other planes diminish, indicating a proper exfoliation procedure. **Figure 2(b)** depicts the atomic force microscopy (AFM) image of the 2D Al QC spin-coated on the Si wafer. **Figure 2(c)** depicts the particle width of the 2D Al QC nanostructure; the average height is 140 nm. **Figure 2(d)** shows the particle height of the 2D Al QC nanostructure; it is found to be 12 nm.

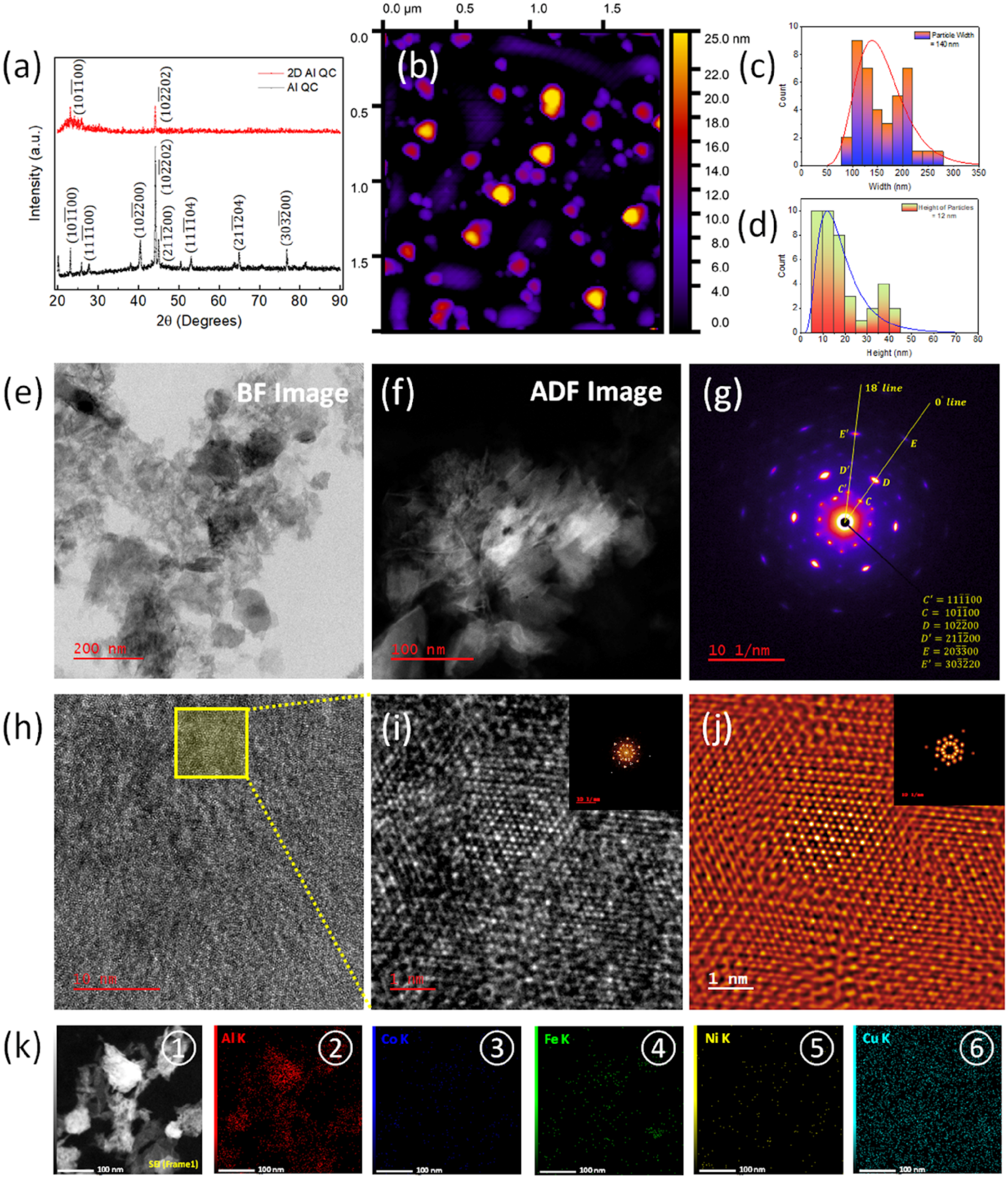

**Figure 2**. (a) X-Ray Diffraction plot for 2D and Bulk Al-QC. (b) Atomic Force Microscopy image of the 2D Al QC on the Si wafer. (c) Width profile of the 2D Al QC dispersed on the Si wafer. (d) Height profile of the 2D Al QC on the Si wafer. (e) Bright Field STEM Image of 2D Al-QC depicting an exfoliated sheet-like nanostructure. (f) Annular Dark Field Image of 2D Al-QC. (g) Main image showing FFT taken from the 2D Al-QC. An inset showing FFT points corresponding to the particular plane. (h) HRTEM Image. (i) Image i corresponds to the zoomed-in view of the selected area in Image h. (j) Inset showing

masked points of the FFT of the HRTEM image 1(i). The main image shows the IFFT of the masked FFT. (k) Image ① shows the SEI image taken in STEM mode. Image shows the ②-⑥ EDS mapping showing the elements in the STEM image.

**Figure 2(e)** shows a 2D Al QC nanostructure on the carbon-coated grid, subjected to STEM imaging. **Figure 2(f)** ADF Image shows defects, different layer-based stacking, and elemental-based compositional variations. The associated SAED pattern demonstrates decagonal symmetry along the 10-fold axis of single-phase decagonal quasicrystals. Liquid exfoliation in IPA yields few-layered nanostructures, as illustrated by **Figure 2(e)**, which are stacked upon one another. The majority of the exfoliated 2D nanostructures of the 2D Al QC are found to be planar sheets, with some nanostructures measuring hundreds of nanometers in lateral width, as confirmed in the AFM. **Figure 2(f)** depicts the ADF image showing different contrast values for different stacking of layered quasicrystal in the grid. The disordered decagonal pattern is repeatable in long-range order and observed in multiple 2D nanostructures. The decagonal SAED pattern shown in **Figure 2(g)**, the SAED pattern derived through the HRTEM mode and was indexed subsequently.[13] Mukhopadhyay et al. developed the least path criterion (LPC) for decagonal quasicrystal indexing based on Fitz Gerald's basis-vector framework.[14] The HRTEM image (**Figure 2(h)**) clearly shows the area in the nanostructure where atoms project the electrostatic potential. The projected atoms manifest as luminous points in the flower-like pattern. Localized overlap between decagonal clusters and the amorphous phase is observed, with contrast variations indicating the collapse of 10-fold symmetry into a completely amorphous structure (**Figure S4c**, Supporting Information). **Figure 2(h)** shows the HRTEM image of the 2D Al QC. Furthermore, the HRTEM and the associated FFT image strongly indicate the presence of decagonal quasicrystalline symmetry in the localized region, as shown in **Figure 2(h)**. **Figure 2(i)** shows a zoomed-in image of the quasicrystalline structure with 10-fold symmetry.[15, 16] The local 10-fold symmetry is evident in the associated fast Fourier transform (FFT) image (**Figure 2(i)**, inset). The quasiperiodic planar atomic configuration is shown in **Figure 2(g)**. The twofold zone-axis diffraction pattern, presented in the inset of **Figure S4d**, is oriented at 90° with respect to the decagonal rotational symmetry axes. The symmetry-breaking characteristic was observed in the SAED pattern of the 2D decagonal nanostructure (Supporting Information **Figure S4d**). A further high-resolution image shows the quasi-crystalline element distribution on the 2D Al QC surface. The quasi-periodic characteristic, with no two-fold symmetry axis, was also noted in the HRTEM, and the associated FFT pattern image is shown in the inset of **Figure 2(i)**.[15] The FFT pattern of the inset of **Figure 2(i)** is masked; further, the IFFT of the masked SAED pattern shows a quasi-crystalline structure with elemental variation, as depicted in **Figure 2(j)**. STEM-based Elemental distribution of the STEM image 2(k) was taken, with a JEOL detector. **Figure 2(k)**① is depicted as the Scanning Electron Image (SEI Image). The elemental uniformity was examined using EDS mapping, which revealed that the alloy 2D nanostructures are uniform, as shown in **Figure 2(k)**②-③-④-⑤-⑥. The SAED pattern confirms the existence of decagonal phase nanostructures exhibiting 2-fold symmetry, see the Supporting Information **Figure S4b**. The 2D decagonal Al QC nanostructures exhibited considerable stability due to the presence of an amorphous phase at the edges, as illustrated in Supporting Information, **Figure S4e**. Supporting Information **Figure S5b** reveals the EDAX spectrum of the 2D Al QC nanostructure. X-Ray Photoelectron Spectroscopy (XPS) results confirm the presence of expected elements in 2D $Al_{70}Co_{10}Fe_{5}Ni_{10}Cu_{5}$ quasicrystal sample (**Figure S1**). Supporting Information Section S1 provides a detailed description of the XPS measurement. As shown in **Figure S2b**, the Al component exhibits the highest intensity, consistent with this 2D Al QC system. The Al 2p region shows a clear metallic Al signature. **Figure S6a** shows two absorption peaks located around 250 nm and 307 nm.[5, 17] The band near 250 nm is predicted to originate from the surface plasmon resonance of the Al component in the 2D Al QC. The band near 307 nm shows a weak absorption feature, which likely arises from the plasmonic activity of other metal components such as Cu, Fe, or Ni.[5, 18] The flattened 2D structure of QC exposes a large fraction of its surface, allowing electromagnetic energy to be localized strongly at the metal surface.

## 3. Results and Discussions

### 3.1 Investigation of Surface Plasmon Resonance Effect in 2D Al QC

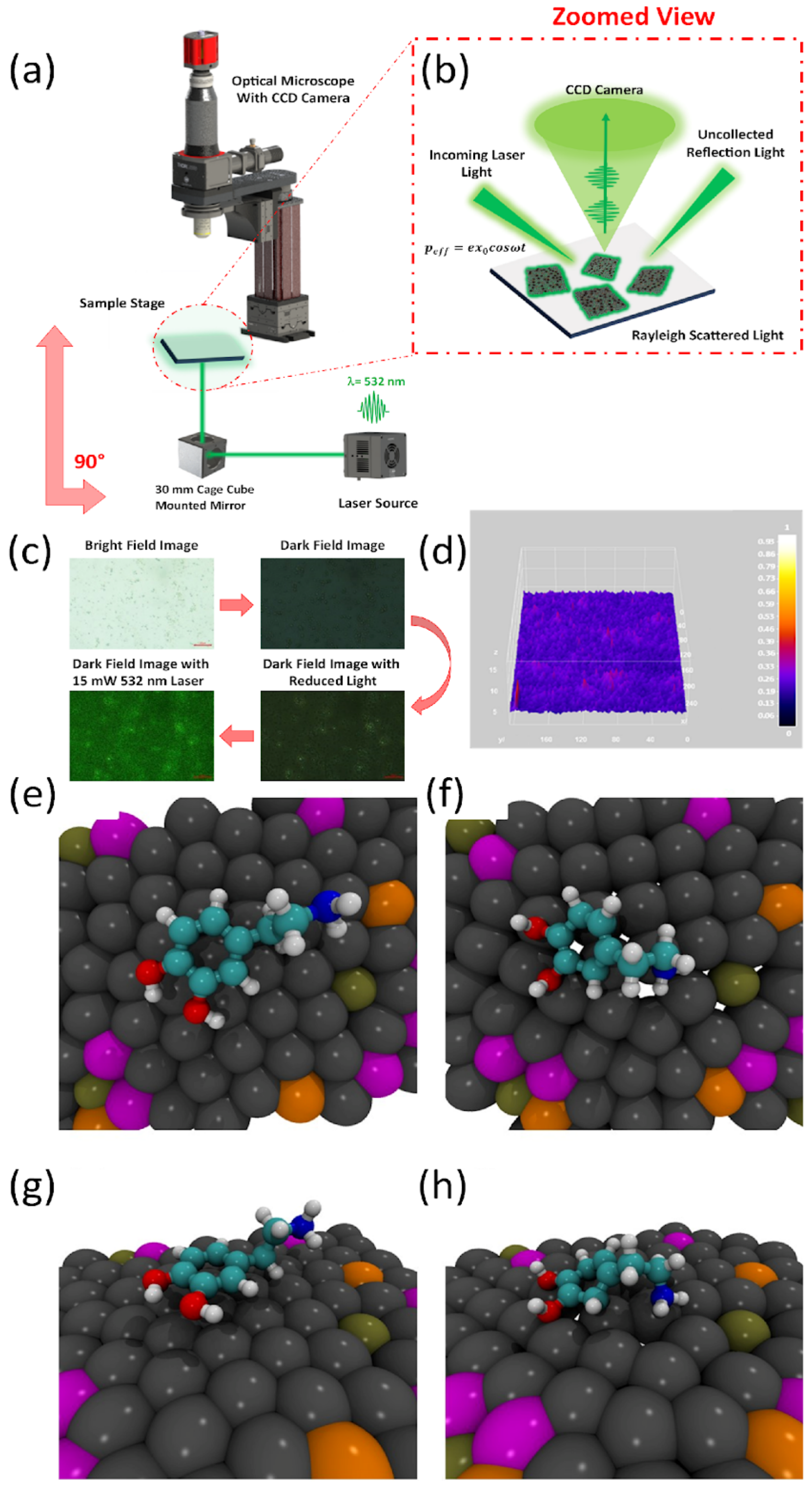

**Figure 3**. Optical setup used to map the surface plasmon resonance of the 2D Al QC. Dark-field optical imaging for verification of the plasmonic structure. (a) Diagram of the optical pathway in the dark-field imaging scheme. (b) Schematic showing the interaction of incident and scattered light at the edge of a 2D nanostructure. (c) Gradual steps taken to visualize the 2D materials with a dark-field approach. Bright-field optical reflection images (left column) Dark Field Image, Dark Field Image with reduced background light, Dark Field Image with 15-mW (λ = 532 nm) illumination. (d) 3D map constructed in ImageJ showing bright particles exhibiting Surface Plasmon Resonance under exposure conditions with 15-mW 532 nm wavelength illumination. (e-f-g-h). Top and side views of the dopamine molecule adsorbed on the quasicrystal (QC) surface, obtained through Monte Carlo sampling and total energy minimization simulations. Panel (e) shows the optimized configuration of dopamine before adsorption, panel (f) depicts the configuration after adsorption on the QC surface, and panel (g) presents a lateral view of the adsorption region. Atoms are represented by spheres of different colors: gray for aluminum (Al), orange for iron (Fe), magenta for cobalt (Co), tan for nickel (Ni), cyan for carbon (C), blue for nitrogen (N), red for oxygen (O), and white for hydrogen (H).

To probe the surface plasmon resonance characteristics of the 2D Al QC, the nanostructure containing IPA is drop-cast onto a glass slide. The 2D Al QC nanostructures were investigated with the transmission mode of the optical microscope. Bright field imaging mode was used to identify the 2D Al QC nanostructure cluster. The glass slide was illuminated with a 15-mW laser beam at 532 nm. Switching the microscope from bright to dark mode revealed scattered light from the 2D Al QC nanostructure. This dark-field mode captures edge-based light scattering, the signature of the SPR effect exhibited by the 2D Al QC nanostructure.[19] The electronic bandgap of the 2D Al QC nanostructure, estimated using the Tauc method, is 1.65 eV. The Aluminium metal exhibits strong plasmonic behavior because it contains a high amount of fixed positive ions and freely moving conduction electrons.[20] The light interacts with the mobile electrons at the surface at an appropriate frequency, and they undergo collective oscillation.[21] Owing to its favorable electronic properties, such as a high electron density and a plasma energy of about 15.6 eV, Al can exhibit pronounced LSPR features similar to those of noble metal nanostructures [22, 23]. **Figure 5(b)** shows that the linear refractive index for 532 nm illumination is approximately 1. As discussed earlier, the Al atoms are primarily responsible for activating the plasma frequency in 2D Al QC.[5] The plasma resonance peak centered near 450 nm deviates from the characteristic of Al value of 532 nm, which can be attributed to the electronic interactions between Al and the other metal constituents in the 2D Al QC lattice. The strong SPR exhibited by the 2D QC is expected to enhance its light-scattering behavior under continuous-wave laser excitation. These scatterings can be evaluated as the scattering coefficient($\sigma$), using the equation, [24]

$$\sigma(\lambda) = \frac{\frac{4\pi^3}{\sqrt{3}}\left(\frac{\eta}{\eta_0} - 1\right)^2 \times x_0^2 \times \left(\frac{\lambda}{x}\right)^{-m}}{Cx^3} \ldots\ldots\ldots\ldots\ldots\ldots\ldots\ldots (1)$$

In this context, x, $\eta$, and C represent the mean length, refractive index, and mass density of the two-dimensional material, respectively, while λ denotes the excitation wavelength. The calibration constant $x_0$ is set to 142 nm, determined empirically. The light scattering behavior of 2D materials is influenced by both, $x$ and the parameter $m$, where $m = 4$ corresponds to small nanosheets applicable under the Rayleigh approximation and $m = 2$ applies to larger nanosheets and aligns with the van der Hulst approximation. For the materials in our study 2D Al QC, the typical value of $x$ is on the order of several micrometers, as found in the AFM results. The estimated scattering cross section σ for the 2D Al QC is approximately $\sim 20 \times 10^3$ $Lg^{-1}m^{-1}$ at an excitation wavelength of 532 nm.[5] Kumbhakar et al. found that the light-scattering response is significantly stronger than that of other 2D materials due to surface plasmon resonance.[5] The discrete nature of the scattered light intensity was further investigated using a model incorporating Rayleigh scattering and charge-dipole interactions.[25-27] Here, the 2D Al QC nanostructure shows enhanced charge-dipole effects, particularly at its edges. Each atom behaves like a polarized sphere carrying charge and an induced dipole moment. At the edges, due to the broken symmetry and altered electronic surroundings compared to the inner regions, these induced dipoles and charges experience stronger local interactions. This leads to increased light scattering at the edges due to Rayleigh mechanisms. The charge-dipole theory explains this phenomenon by treating atoms as light-responsive dipoles. Edge atoms, missing neighbors on one side, develop stronger polarization, intensifying their scattering contribution relative to atoms within the sheet's bulk. This edge-specific enhancement in scattering is a direct consequence of the distinct charge and dipole distributions induced by finite size and edge boundary conditions in two-dimensional materials such as graphene. The effective electric dipole vector at the edge is more prominent compared to that of the edge to that of the bulk, and the expression is described as,

$$p_{eff} = ex_0 cos\omega t \ldots\ldots\ldots (2)$$

Here, $e$ represents the charge of a unit electron, $x_0$ is quantified as the oscillation amplitude of the dipole, and ω is the angular frequency of the incident light. Rayleigh scattering occurs when incident light interacts with dipoles and is scattered. This Rayleigh scattering ($\bar{I}$) is given by the expression,[28]

$$\bar{I} = \overline{I_0} \frac{c\pi^2 |p_{eff}|^2}{2\varepsilon_0 r^2 \lambda^4} (1 + cos^2\theta) \ \ldots\ldots\ldots (3)$$

In this equation, the $\overline{I_0}$ is the average intensity of incoming light, *c* is the speed of light in the vacuum, $\theta$ is the angle between the optical axis of the objective lens and the incident direction of the incoming light, $\varepsilon_0$ is the permittivity of the vacuum, $\lambda$ is the wavelength of the incoming light, and *r* is the distance between the objective lens and the electric dipoles. It is confirmed that the scattering follows a second power law with $p_{eff}$, as $\ddot{p}_{eff} = \quad {}^{2} p_{eff}$. For 2D materials, in plane interactions are much weaker than edge scattering. Hence, the total Rayleigh scattering can be considered expressed as a sum of scattering from individual layers.

Here, dark-field optical microscopy was used to detect this edge-based light-scattering effect as shown in **Figure 3(a)**. This method can be implemented to recognize even a single layer of a graphene flake.[29] The darkfield approach relies on scattered light rather than reflected light and can be achieved with various substrates. The darkfield approach is highly dependent on factors related to edge scattering, such as the strength of the dipoles at the edge. **Figure 3(b)** depicts the edge scattering collected by the CCD camera. This edge scattering depends on the surface-plasmon behavior of the 2D Al QC. This surface plasmon behavior enhances edge scattering, as seen in the dark-field image, **Figure 3(c)**, illuminated by a 532 nm laser. These properties enable a new, highly sensitive method for detecting foreign molecules, which can be combined with 2D Al QC. Different ppb solutions of dopamine (dopamine hydrochloride, 99%, Sigma-Aldrich) were mixed with a 2D Al QC solution; these dopamine molecules interact with the hindered surface plasmon resonance of the 2D Al QC molecules. A study was performed to determine whether the interaction between dopamine and 2D Al QC.

### 3.2 Computational Details

The two-dimensional quasicrystal (QC) structure 2D $Al_{70}Co_{10}Fe_5Ni_{10}Cu_5$ was investigated using electronic structure simulations performed with the density functional tight-binding (DFTB) method[30], employing the Periodic Table Baseline Parameter (PTBP) Slater-Koster set[31] to describe atomic interactions, through the DFTB+ software.[32] To better mimic the experimental conditions, solvation effects were included through the Generalized Born Surface Area (GBSA) approach [33] with the dielectric parameters for isopropanol (IPA). For both the QC and the dopamine molecule, the convergence thresholds were established at $10^{-5}$ Ha for total energy and self-consistent charge (SCC) iterations, and $10^{-4}$ Ha/Bohr for the maximum force component during geometry optimization, respectively. The simulations were carried out with k-point sampling restricted to the Γ point. The QC lattice was relaxed in plane while constraining the z-axis to maintain its structural integrity, and the dopamine molecule was fully optimized. To explore adsorption preferences, the dopamine-QC interaction was analyzed using a Monte Carlo algorithm, in which a configuration space of approximately 250,000 adsorption geometries was screened, and the two lowest-energy configurations were selected for subsequent refinement. These energetically preferred sites were consistently associated with surface regions exhibiting higher aluminum density.

Following this screening, we refined the vertical separation between the dopamine molecule and the QC surface by systematically varying the interfacial distance by ±0.20 Å around the adsorption minimum. The binding energy ($E_b$) was computed using the equation,

$$E_b = E_{QC+Mol} - \left(E_{QC} + E_{Mol}\right) \ldots\ldots\ldots (4)$$

Here, $E_{QC+Mol}$ represents the total energy of the dopamine-C complex, $E_{QC}$ is the total energy of the isolated quasicrystal, and $E_{Mol}$ corresponds to the energy of the isolated dopamine molecule. To reduce computational cost while preserving the local environment of interest, a structural fragment centered at the most stable adsorption site was extracted from the optimized model and subsequently reoptimized. The original QC supercell contains 352 atoms, while dopamine consists of 22 atoms, leading to 384 atoms in the full system before fragment reduction. Finally, to gain further insight into the structural and electronic modifications induced by adsorption, the Raman and infrared spectra of dopamine were simulated, both before and after adsorption on the quasicrystal surface. These results were then compared with available experimental spectra to elucidate changes in molecular vibrational modes associated with the adsorption process. The experimental results revealed that dopamine undergoes significant changes in its optical absorption characteristics and other physicochemical properties upon interaction with the quasicrystal surface. These changes suggest that the adsorption process affects not only the structural conformation of dopamine but also its electronic distribution. Driven by these findings, we performed further computer simulations to explore the structural and chemical changes induced by adsorption, to elucidate the molecular mechanisms responsible for the observed experimental behavior. As previously discussed, the initial adsorption analysis carried out using a Monte Carlo-based sampling algorithm indicated that the most energetically favorable adsorption sites were in regions with the highest aluminum concentration, rather than in areas containing a mixed distribution of metals (see **Figure 3(e-f-g-h)**). Based on these observations, we carried out simulations to investigate the interaction mechanisms between the quasicrystal surface and the dopamine molecule in detail. The most energetically favorable adsorption sites for dopamine were identified, and a total-energy minimization was performed along the z-direction to obtain the equilibrium configuration. The calculated binding energy for the optimized system was -3.13 eV, with a vertical separation of 2.28 Å from the OH side and 2.08 Å from the $-NH_2$ side, between the dopamine molecule and the quasicrystal surface (see **Figure 3(e-f-g-h)**). These strongly negative binding energies and short adsorption distances indicate a robust interaction between dopamine and the quasicrystal surface, suggesting considerable electronic coupling and potential structural rearrangements at the adsorption interface. To gain further insights into the interaction mechanisms, a representative fragment of the dopamine–quasicrystal (QC) system was isolated and further optimized (see **Figure 3(e-f-g-h**) and **Video 1**). After structural relaxation, the dopamine molecule remained intact, in contrast to the typical molecular dissociation observed for species adsorbed on highly reactive metallic surfaces.[34] In this case, the hydroxyl (-OH) group forms a coordination bond with an aluminum atom on the QC surface, while the amine ($-NH_2$) group undergoes a rotation of approximately 180°, establishing a hydrogen-bridged interaction oriented toward the aluminum site. This configuration suggests that the QC surface does not promote molecular degradation but instead stabilizes dopamine through specific interfacial bonding. The molecule thus preserves its overall morphology, undergoing significant conformational changes that optimize its electronic and geometric compatibility with the quasicrystal surface.

### 3.3 Nonlinear Optical Properties-based Sensing

The high-amplitude sonication vibrations cause the layered sheets to exfoliate into a few layers of two-dimensional nanostructure. The concentration of 2D Al QC material in the IPA solution was 14 $mg.mL^{-1}$. Rhodamine B dye was mixed with the solution at a lower concentration of 0.0025 $mg.mL^{-1}$, as the 2D Al QC has a plasmonic nature. Incorporation of a gain medium into the system can enhance absorption of the incident laser beam and provide a more favorable platform for dopamine interaction, owing to the resulting local-field enhancement.[35] Rhodamine B was selected as the dye due to its strong absorption at 532 nm. A 1100 ppb dopamine solution was used as a reference solvent to mix with the solution. After confirming the interaction between the material and dopamine, UV-Vis absorption spectra of the prepared mixtures were used to assess the material's ability to detect dopamine. 2.5 mL of the prepared material solution and 0.5 mL of the dye (RhB) were placed in a standard 3.5 mL cuvette, and a dopamine solution at 1100 ppb was added at various concentrations, as shown in Table 2, using a micropipette.

**Table 1. Concentration of 1100 ppb Dopamine solution mixed of Dopamine during Laser Impact:**

| Steps taken to mix 1100 ppb Dopamine solution to 2D-Al-QC@RhB system | Concentration of 1100 ppb Dopamine solution mixed (mL) |
|---|---|
| 1 | 0.5 |
| 2 | 1 |
| 3 | 2.5 |
| 4 | 4 |
| 5 | 5.5 |

To probe the interaction of the 2D-Al-QC@RhB system with dopamine, 1100 ppb of dopamine was added to the system, and the diffraction rings were observed. The rings in the far-field diffraction pattern started to get quenched as soon as the dopamine was added, and a complete null diffraction pattern was observed after a certain level of dopamine solution was added. After the addition of dopamine, the observed time evolution increases. The corresponding glowing colloids in the cuvette for the 2D-Al QC and 2D-Al QC@RhB+1100 ppb. The ring patterns are destroyed, so any changes in the patterns used to sense dopamine concentration are not viable.

However, the interaction time between 2D-Al QC and dopamine can be successfully recorded using this method. Given the negligible dye concentration, the possibility of dye degradation is ruled out. Initially, the nonlinear refractive index of the only dye solution with 2D Al QC is calculated. The optical configuration for spatial self-phase modulation (SSPM) spectroscopy is depicted in **Figure 4(a)**. This approach allows the extraction of key nonlinear optical parameters, including the nonlinear refractive index ($n_2$) and third-order susceptibility ($\chi^{(3)}_{total}$), of 2D Al QC suspended in solution. Three CW lasers at 650, 532, and 405 nm, housed in Thorlabs temperature-controlled Ø5.6 mm diode mounts, were used for the measurements. Laser beams were focused through a Thorlabs N-BK7 Plano-Convex Lens (focal length of 20 cm) onto the cuvette containing the sample. The SSPM effect is observed as the focused laser beam passes through the cuvette, generating a far-field diffraction profile that is recorded using a CCD camera. The nonlinear optical response of the 2D material in this process is governed by Kerr nonlinearity, which leads to radial redistribution of the beam intensity and the formation of the diffraction pattern. Rayleigh scattering is negligible in this case. Although thermal interaction causes the central part of the diffraction profile to be dilated. The total refractive index can be defined as,

$$n(r) = n_0 + n_2 I(r) \ \ldots\ldots\ldots\ (5)$$

Here, $n_0$ depicts the linear refractive index, while I indicate the intensity of the incoming laser. The expression of the nonlinear refractive index $n_2$ is found to be:

$$n_2 = \left(\frac{}{2n_0 L_{eff}}\right) . \frac{dN}{dI} \ \ldots\ldots\ldots\ (6)$$

The parameter $\frac{dN}{dI}$ is defined as the variation in ring number with intensity and is used to determine the nonlinear refractive index of 2D-Al-QC@RhB. The $\chi^{(3)}_{total}$ is used to predict the nonlinear optical properties of materials, providing insight into how strongly a material's electronic cloud distorts under strong optical fields and revealing information about electronic structure, bandgap, excitonic effects, and polarizability. The $\chi^{(3)}_{total}$ expressed as,

$$\chi^{(3)}_{total} = \frac{cn_0^2}{12\ ^2} 10^{-7} n_2\,(e.s.u) \ \ldots\ldots\ldots\ (7)$$

Here, the term c represents the speed of light in a vacuum, $n_0$ denotes the linear refractive index of the IPA solvent, and $n_2$indicates the effective length that the laser beam traverses within the cuvette. The amount of material available in the solution accounts for the value of $\chi^{(3)}_{total}$ defined above.

For different operating conditions (such as wavelength, concentration of 2D Al QC, and cuvette width)[7], the value of $\chi^{(3)}_{total}$ might change. Also, for different concentrations of dopamine solutions, the value of $\chi^{(3)}_{total}$ might differ. Here we investigate the $\chi^{(3)}_{monolayer}$, the effective third-order nonlinear susceptibility for one layer of 2D Al QC. The relation between $\chi^{(3)}_{total}$ and $\chi^{(3)}_{monolayer}$ can be defined as,

$$\chi^{(3)}_{total} = N^2_{eff}\chi^{(3)}_{monolayer} \ \ldots\ldots\ldots\ (8)$$

**Table. 2 Concentration of Dopamine solution mixed dependent nonlinear refractive index $n_2$, Third-order nonlinear susceptibility $\chi^{(3)}_{total}$, and Third-order nonlinear susceptibility for a monolayer $\chi^{(3)}_{monolayer}$.**

| Volume of 1100 ppb Dopamine Concentration | $dN/dI$ $(Wcm^{-2})$ | $n_2$ $(cm^2W^{-1})$ | $\chi^{(3)}_{total}$ (e.s.u.) | $\chi^{(3)}_{monolayer}$ (e.s.u.) |
|---|---|---|---|---|
| $0\ mL$ | $2.3$ | $4.4539710^{-5}$ | $0.00215$ | $2.4336210^{-12}$ |
| $0.5\ mL$ | $1.69$ | $3.272710^{-5}$ | $0.00158$ | $1.7881810^{-12}$ |

| $1\ mL$ | $0.77$ | $1.4911110^{-5}$ | $7.1929810^{-4}$ | $8.1473210^{-13}$ |
|---|---|---|---|---|
| $2.5\ mL$ | $0.61$ | $1.1812710^{-5}$ | $5.6983410^{-4}$ | $6.4543710^{-13}$ |

Here, $N_{eff}$ is the effective number of 2D Al QC layers. Sensitivity is calculated to determine whether the plasmonically enhanced 2D Al QC@RhB system can detect a small change in dopamine concentration. The 1100 ppb solution is mixed with the cuvette in 0 mL, 0.5 mL, 1mL, and 2.5 mL. For each concentration, the nonlinear refractive index is found to be 4.459 $\times 10^{-5}$, $3.27\times 10^{-5}$, $1.459\times 10^{-5}$, and $1.18\times 10^{-5}$ $cm^2W^{-1}$. Figure 4(b) shows the increment in the number of rings with an increase in the laser beam ($\lambda$= 532 nm) intensity.

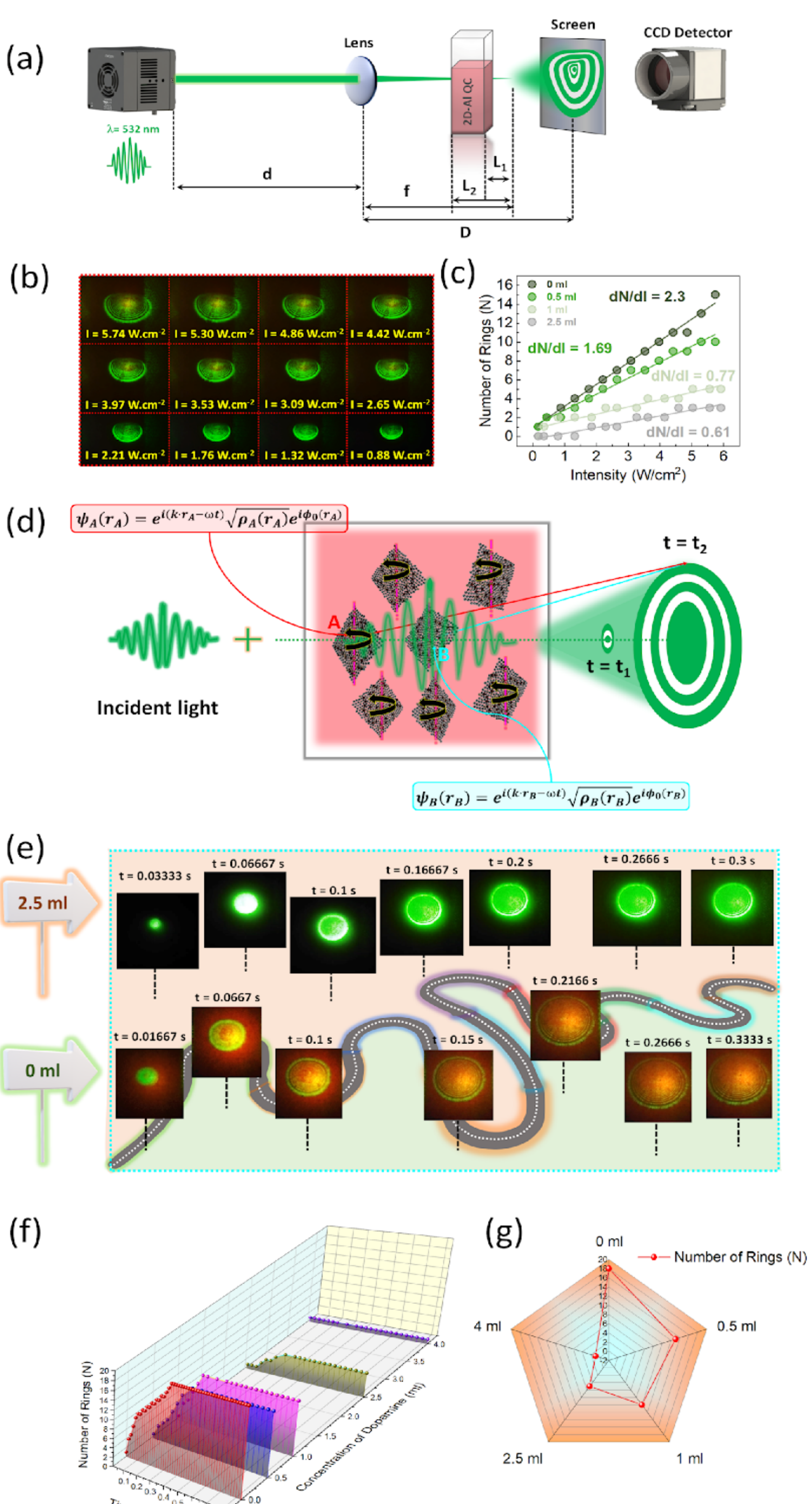

**Figure 4**. (a) SSPM Spectroscopy setup. (b) Generation of diffraction profile on the far screen with variation in intensity for wavelength λ = 532 nm for 2D Al QC-RhB system. (c) $dN/dI$ value for different mL of dopamine mixing with the 2D Al QC-RhB system. (d) Depiction of the "Windchime model" and full diameter diffraction pattern formation on the far screen due to enhanced electronic coherence from time to time $t = t_1$ to $t = t_2$. (e) Temporal progression of SSPM patterns across 1100 ppb dopamine solution mixing of 0 mL and 2.5 mL at a fixed wavelength of λ= 532 nm and intensity on the 2D Al QC-RhB system. (f) Response time under the influence of different concentrations of dopamine mixed in liquid form. (g) Generated number of Rings under the influence of varying levels of dopamine solution exposure.

As the intensity increases due to the SSPM effect, a greater phase shift is introduced, leading to a larger number of rings in the diffraction profile. Adding a 1100 ppb solution at different levels reduces the concentration of active 2D Al QC nanostructures. The solution is mixed in a 2D Al QC@RhB container to 80 ml and sonicated to make a homogeneous solution. The dopamine molecule binds to the 2D Al QC and reduces polarization under laser illumination. An increment of this dopamine solution in the cuvette diminishes the SSPM effect, causing a reduction in the number of SSPM rings. This decrement in SSPM rings under the same laser intensity illumination causes the $\frac{dN}{dI}$ value to decrease, with the increase in 1100 ppb dopamine addition. The value of $\frac{dN}{dI}$ is investigated to be 2.3, 1.69, 0.77, and 0.61 cm$^2$W$^{-1}$ for 0, 0.5, 1, and 2.5 mL of 1100 ppb of dopamine solution addition, as shown in Figure 4(c). The sensitivity formula is defined as,

$$Sensitivity = \frac{Change\ in\ Ouput}{Change\ in\ Input} \ldots\ldots\ldots \quad (9)$$

Or,

$$Sensitivity = \frac{Change\ in\ nonlinear\ Refractive\ Index}{Change\ in\ Active\ Dopamine\ Solution} \ldots\ldots\ldots \quad (10)$$

The sensitivities are 2.378, 3.56, and 0.62 cm$^2$W$^{-1}$mL$^{-1}$ for the changes at each step of dopamine solution mixing. Hence, it is found that the maximum refractive index-based sensitivity shown by the 2D Al QC@RhB system for sensing the Dopamine molecule through the SSPM Spectroscopy method is 3.56 cm$^2$W$^{-1}$mL$^{-1}$. As a further pictorial analysis method, time-evolution-based sensitivity is evaluated.

### 3.4 Wind Chime Model: Diffraction Pattern Generation based on Dopamine Sensing under 532 nm wavelength illumination

This work examines pattern generation and the SSPM mechanism. Initially, the 2D Al-QC nanostructures are randomly oriented within the solvent; initially, the interaction between the incoming laser beam and the nanostructure is minimal. Over time, a diffraction pattern begins to manifest on the distant screen. Wu et al. developed a model to clarify the formation of SSPM patterns, including the time necessary for a constant intensity.[36] The proposed theory assumes that the first contact between the incident laser beam and the 2D Al-QC takes place at random angles relative to the direction of the laser's electric field. The solution's hanging two-dimensional Al-QC is polarized by the incoming laser beam.[24] Energy relaxation facilitates the alignment of polarized 2D Al-QC with the incoming beam's electric field and the polarized axis of the 2D Al-QC.[36] The hydrodynamic angle between the 2D Al QC nanostructure and the external field diminishes over time, aligning more closely with the electric field and increasing the ring intensity in the diffraction profile. The maximum number of rings is observed when complete alignment of all 2D Al-QC occurs along the laser beam's polarization axis. We subsequently investigate the development of ac electron coherence. Expressing the electronic wave function at $r_A$ as $\psi_A(r_A) = \sqrt{\rho_A(r_A)}e^{i\phi(r_A)}$, with denoting the local electron density $\psi^*_A(r_A) \cdot \psi_A(r_A) = \rho_A(r_A)$ and the phase $\phi(r_A) = k \cdot r_A - \omega t + \phi_0(r_A)$ of the electronic wave function, that is entirely dictated by the extrinsic light field (Figure 4(d)). This denotes an enforced oscillation, in which electrons are required to conform to the local phase established by the extrinsic field, even in the presence of scattering and interactions. In this context, $\phi_0(r_A)$ represents the phase lag as a reaction to the extrinsic field. The initial arbitrary phase is intact. The same idea applies to the electrical wave functions at $r_B$, regardless of whether they are different or the same nanostructure. The two wave functions are coherent with the incoming laser pulse and therefore are coherent with one another. This relates to dynamic or ac electron coherence, in contrast to the widely recognized steady-state or dc electron coherence in transport measurements. The electrons stay near their equilibrium positions. In accordance with the wind-chime configuration, electron coherence increases as more electrons become correlated, while the collision frequency decreases due to the perfectly oriented shape. The intensity at the screen is $I(r_C) = E^*(r_C) \cdot E(r_C)$, here $E(r_C)$ is denoted as the electric field of the laser beam at $r_C$. The electric field becomes $E(r_C) = \varsigma(\chi^{(3)})\Psi(r_C)$, with phase being $\Psi(r_C) = e^{i[k\cdot\tau_C - \omega t + \phi_0]}\left[\sqrt{\rho_A}(r_A)e^{-ik_1\cdot\tau_A} + \sqrt{\rho_B}(r_B)e^{-ik_1\cdot\tau_B}\right]$, here the wave vector $k_\perp = d\Delta\phi/dr$ is included due to Kerr nonlinearity. (Figure 4(d)) depicts the influence of external light field on the electrical wave function. We investigated the influence of the strong laser beam's polarization on suspended 2D Al-QC in IPA to confirm this model,

$$N = N_{max}\left(1 - e^{\frac{t}{\tau_{rise}}}\right) \ldots\ldots\ldots (11)$$

**Figure 4(e)** illustrate the temporal evolution of the diffraction pattern for wavelengths ($\lambda$ = 532 nm) at dopamine mixing quantities of 0 mL and 2.5 mL, respectively. **Figure 4(d)** depicts the time evolution for different levels of dopamine mixing. The respective laser beam intensities are 6.28 $W.cm^{-2}$, which have been set constant throughout the experiment. Supporting Information **Figure S10** shows the temporal evolution of the rings at varying dopamine concentrations. The maximum number of rings is 18, 13, 8, 5, and 1 for different levels of dopamine mixing concentration, with an incoming laser beam intensity of 6.28 $W.cm^{-2}$. Here, $N$ corresponds to the number of rings appearing in the diffraction profile at a particular instant, $N_{max}$ indicates the maximum quantity of rings generated under constant laser intensity, and $\tau_{rise}$ denotes the duration required for pattern formation. The number of rings at 532 nm for 0 mL exceeds that of other concentrations, signifying augmented light-matter interaction, which results in an extended duration to produce diffraction patterns relative to those having different dopamine concentrations. The temporal evolution of the diffraction profile, leading to the maximum number of rings, is presented in **Figure 4(e)** for various dopamine concentrations. The dopamine molecules bind to the 2D Al QC, hindering its electronic polarization and thereby suppressing ring generation in the diffraction profile. The presence of dopamine in the 2D Al QC further degrades electronic behavior and optical excitation. Hence, this SSPM method can be used to detect the dopamine molecule and the acute concentration of the material mixed with the 2D Al QC solution. The time evolution-based sensitivity is defined as,

$$Sensitivity = \frac{Change\ in\ Ouput}{Change\ in\ Input} \ldots\ldots\ldots\ (12)$$

Or,

$$Sensitivity = \frac{Change\ in\ Maximum\ Number\ of\ Rings\ (N_{max})}{Change\ in\ Active\ Dopamine\ Solution} \ldots\ldots\ldots\ (13)$$

The sensitivities are found to be 10 $mL^{-1}$, 10 $mL^{-1}$, and 2 $mL^{-1}$ for the change in each step of dopamine solution mixing. Hence, it is found that the time-evolution-based sensitivity of the 2D Al QC@RhB system for sensing the Dopamine molecule using the SSPM Spectroscopy method is 10 $mL^{-1}$.

### 3.5 Traditional Spectroscopic Methods-Based Sensing:

**Figure 5(a)** shows the absorbance spectra of the 2D AL QC, before and after mixing the dopamine. The absorbance spectrum of 2D-Al QC has a principal peak at 250 nm and a secondary peak at 307 nm. From the UV-Vis absorbance spectrum, the linear absorbance coefficient can be calculated. This relationship can be explained as,

$$\alpha(\omega) = \frac{2.0303A(\omega)}{d} \ldots\ldots\ldots\ (14)$$

The investigated absorption coefficients (**Figure S11a**) show an intense match throughout the visible spectrum. Interestingly, any mismatch in the infrared region is minimal, and given that typical absorption is close to $10^5$ $cm^{-1}$, its practical effect seems negligible. Such slight differences are likely due to effects from the sample's substrate or perhaps some minor impurity present on the surface during measurement, rather than an intrinsic property of the material itself. **Figure 5(a)** highlights a distinct peak that can be attributed to the plasmon resonance frequency, which drives the optical nonlinearities observed in two-dimensional quasicrystals. As detailed in the experimental section, it is primarily the aluminum atoms that initiate this plasma frequency within the 2D QC matrix. Notably, the plasmon resonance peak is observed around 450 nm, remarkably close to the known value for aluminum at 532 nm laser illumination. Where d = sample thickness; here, the sample thickness is set to 1 cm, the same as the cuvette thickness loaded with the 2D material. The extinction coefficient was calculated. The real refractive index can be calculated from the Kramers-Kronig relations applied to $k(\omega)$. A MATLAB-based script was developed to calculate the linear refractive index from the extinction coefficient, obtained from the absorbance spectrum of the 2D material-containing cuvette.

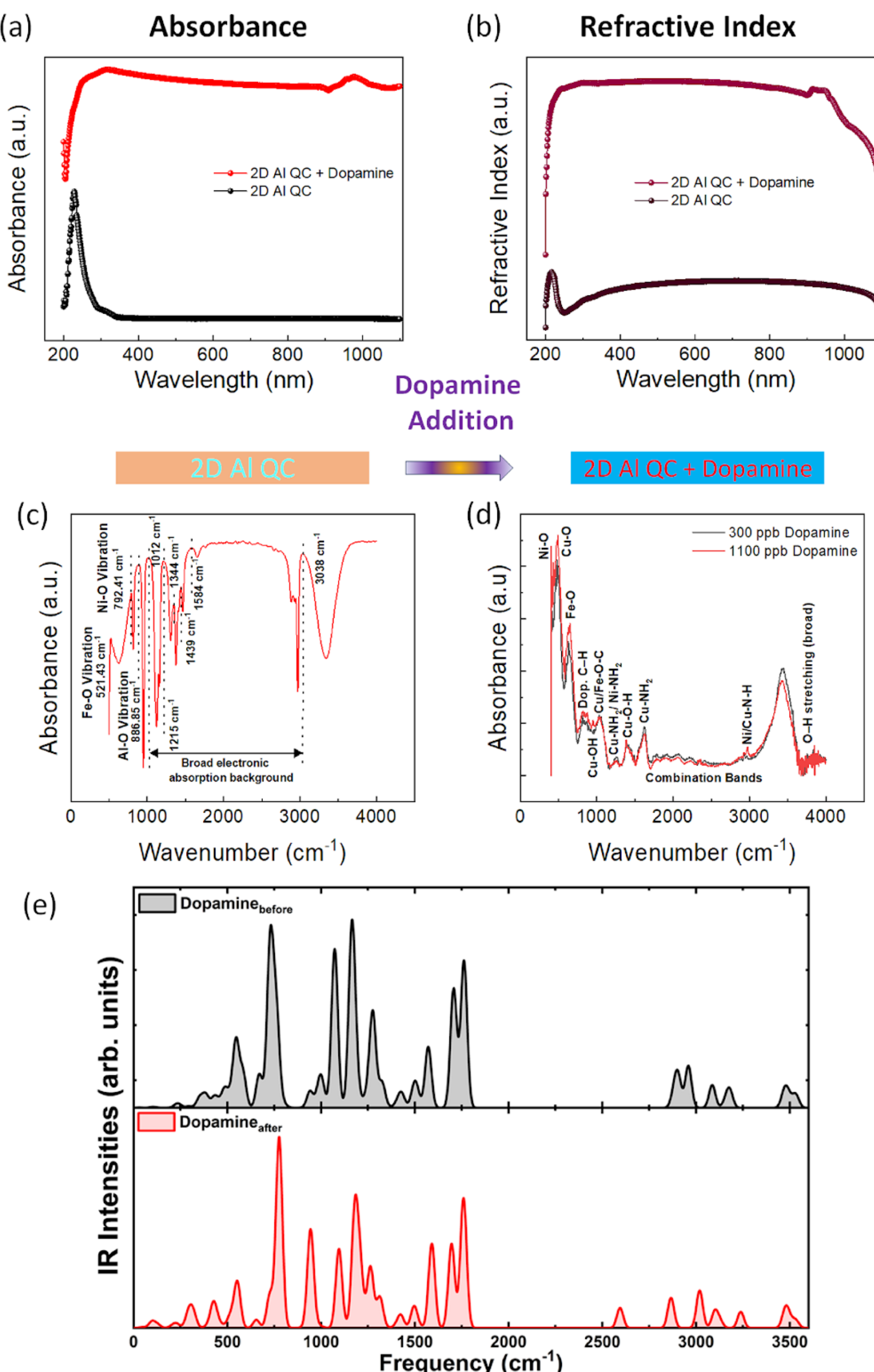

**Figure 5**. (a) Absorbance data, and (b) Refractive index from the UV-vis Spectroscopy before and after 2D Al-QC-Dopamine interaction. (c-d) Infrared spectra of dopamine before and after adsorption on the quasicrystal (QC) surface. (e) The comparison highlights the vibrational mode shifts and intensity variations associated with structural and electronic changes induced by adsorption.

From this relationship, the linear refractive index can be calculated. Hence, **Figure 5(a-b)** shows the comparison of absorbance and the linear refractive index of the 2D Al QC and 2D Al QC with dopamine mixed, respectively. In both cases above, a change in the spectrum is observed. The change in the nature of the linear refractive index (Supporting Information **Figure S11a** and **S11b** shows the absorption and extinction coefficients of the 2D Al QC and the 2D Al QC with dopamine mixed, respectively. The 1100 ppb dopamine was mixed with 2D Al QC, and the absorbance spectrum was re-investigated. The absorption coefficient, extinction coefficient, and linear refractive index were recalculated to assess the effect of dopamine on the 2D Al QC. It was shown that dopamine mixing alters the nature of the above-mentioned parameters. Hence, the portion of the dopamine mixing can be calculated. Following the addition of dopamine, 5 minutes of active bath sonication were performed to facilitate interaction. The principal peak at 307 nm exhibits a red shift, whereas the secondary peak at 250 nm gradually diminishes with rising dopamine concentration. The secondary peak is entirely suppressed upon mixing at a dopamine concentration of 1100 ppb. This suggests that the secondary peak arises from localized surface plasmon resonance (LSPR). As dopamine molecules interact with surface and edge atoms, this effect gradually diminishes, as evidenced by the attenuation of the peak. The principal peak at 307 nm, however, intensifies and redshifts with increasing dopamine levels. This is shown in Supporting Information **Figure S7a**. The redshift was at least 4 nm at a dopamine concentration of 400 ppb (about 19.5 nM). Below this concentration, the peak shift is minimal, whereas beyond it, both the peak shift and intensity increase linearly. Therefore, the linear detection range for dopamine using this approach is 300-700-1100 ppb (about 19.5-71.5 nM). At a dopamine concentration of 1100 ppb, the main peak exhibits a shift of 11 nm, occurring at 307 nm. At concentrations approximately 1100 ppb, the fast interaction of dopamine with the surface and edge Al (Aluminium) atoms of the 2D-Al nanostructure via its -OH group results in the formation of Al-O and Zr-O bonds, leading to the agglomeration and precipitation of the newly synthesized material. In Supporting Information **Figure S7b**, the principal absorbance peak for 2D-Al QC at 307 nm exhibits a total redshift of 11 nm upon the addition of dopamine in concentrations between 0 and 1100 ppb, following a 5-minute sonication period of the solution. After 30 minutes of resting the solution and a 5-minute follow-up bath sonication, the UV-Vis Spectroscopy measurement was repeated for all solutions. All corresponding peaks were found to be quenched. Hence, no shift was observed for the 307 nm peak, as illustrated in the Supporting Information **Figure S7d**. **Figure 5(c)** shows the FTIR spectrum of the pristine 2D Al QC, which exhibits characteristic spectra of Al-O stretching bands at 485 cm$^{-1}$, 570 cm$^{-1}$, and 820 cm$^{-1}$.[37, 38] The peak at 627 cm$^{-1}$ was found to correspond to Cu-O bond stretching.

### 3.6 FTIR Examination of a 2D Al QC nanostructure with Dopamine Functionalization

The FTIR spectrum of the dopamine-functionalized 2D Al QC, consisting of $Al_{70}Co_{10}Fe_5Ni_{10}Cu_5$, displays unique vibrational characteristics that clarify the bonding environment and the interaction between dopamine and the metal-rich surface. The most pronounced absorption band, occurring between 500 and 550 cm$^{-1}$, is attributed to the stretching of metal-oxygen (M-O) bonds. Aluminum ($Al^{3+}$) exhibits significant oxophilicity and fully utilizes the Al-O vibrations in this context. The Fe-O and Cu-O bond lengths may contribute, as iron and copper are known to form oxides or hydroxides on surfaces upon exposure to air or water. Nickel and cobalt may expand this property by superimposing Ni-O and Co-O modes. The feeble absorption bands between 800 and 1000 cm$^{-1}$ result from C-H bending vibrations occurring out of the plane of the aromatic ring in dopamine. Interactions between metal-OH and metal-ligand, particularly in the presence of catechol or amine coordination, can also induce these effects. Small peaks in the 1200-1500 cm$^{-1}$ range correspond to C-N stretching from the primary amine of dopamine and C=C stretching from the aromatic ring. A broader band at 1600 cm$^{-1}$ is associated with N-H bending and aromatic ring vibrations. A pronounced absorption band exists between 3400 and 3500 cm$^{-1}$, indicating O-H stretching vibrations from the phenolic hydroxyl groups of dopamine and N-H stretching. This expansion suggests significant hydrogen bonding, maybe resulting from interactions with the quasicrystal surface, hydroxylated metal sites, or adsorbed water molecules. The FTIR measurements indicate that dopamine successfully functionalized the entire exfoliated quasicrystal. The principal interactions appear to be between the surface oxides Al-O and Fe-O, as well as between the catechol and amine groups of dopamine and the metal ions $Cu^{2+}$, $Fe^{3+}$, and $Ni^{2+}$. The formation of metal-ligand complexes broadens and shifts the low-wavenumber region. The integrated spectral characteristics indicate the formation of a hybrid metal-organic interface between the dopamine molecules and the quasicrystal's surface.

FTIR analysis of **Figure 5(d)** shows the interaction of 2D Al QC with an equal amount of dopamine interaction of different concentrations, 300 ppb and 1100 ppb, respectively. FTIR Analysis of Dopamine-Functionalized 2D QC at Different Concentrations. The FTIR spectra of the dopamine-functionalized 2D QC ($Al_{70}Co_{10}Fe_{5}Ni_{10}Cu_{5}$) at concentrations of 300 ppb and 1100 ppb exhibit significant variations indicative of dopamine adsorption and its interaction with the metallic surface. Significant alterations occur in the low-wavenumber region (about 500-800 $cm^{-1}$). For instance, at 1100 ppb, the absorbance generally increases. This region is typically associated with metal-oxygen (M-O) stretching vibrations, and increased intensity indicates greater surface coordination between dopamine and metal sites. The spectra exhibit minor alterations in the mid-infrared range (1200-1500 $cm^{-1}$), attributable to the C-N stretching vibrations of the amine group and the C=C skeletal vibrations of dopamine. These characteristics become more pronounced at elevated doses, indicating an increased presence of dopamine on the surface. The O-H and N-H stretching vibrations produce a prominent and intense absorption band within the range of 3400 to 3500 $cm^{-1}$. This band exhibits significant broadening and enhancement at 1100 ppb, indicating an increased presence of phenolic hydroxyl and amine groups that are engaging in hydrogen bonding with the quasicrystal surface or interacting with dopamine. The formation of the concentration-dependent spectrum indicates effective functionalization, characterized by intensified and broader metal-organic interactions with elevated dopamine levels. The notable absorption in the 500-550 $cm^{-1}$ range is mostly ascribed to Al-O stretching, owing to aluminum's pronounced oxophilicity and its tendency to form surface oxides, particularly after exfoliation in polar solvents. The elevation in this region with elevated dopamine levels may indicate that the Al-O bonds are either more robust or more abundant. This may be due to dopamine coordinating with them. The increased absorbance intensity and broadening of the FTIR bands with increasing dopamine levels unequivocally indicate that dopamine exhibits a higher binding affinity for the surface metal sites of the quasicrystal. Aluminum and iron primarily engage in oxide bonding and catechol chelation, respectively, whereas copper and nickel facilitate further complexation via interactions with amine and hydroxyl groups. The observed vibrational changes upon dopamine adsorption, particularly the emergence and splitting of specific Raman and infrared bands, suggest that the quasicrystal surface is highly sensitive to local electronic and structural perturbations induced by molecular binding. Such distinct spectroscopic fingerprints can serve as detectable signals for the presence of dopamine, indicating that the QC has potential applicability as a molecular sensor. Comparing both spectra depicting different levels of interaction of 300 and 1100 ppb with 2D Al QC, as shown in **Figure 5(d)**. It was concluded that the most distinguished/enhanced signal derives from the region 750 $cm^{-1}$. This is most likely due to the vibrational component of Al-O; the signal is dominant because of surface alumina or hydroxylation, and Ni-O may also contribute. The most characteristic peak is at 935 $cm^{-1}$. It is concluded that this peak shift is observed due to the interaction of dopamine molecules with the 2D Al QC. A peak shift in the major peak at 820 $cm^{-1}$ was observed; Al-O bonding contributes primarily through this process. A peak shift in the other major peaks was also observed.[39] With respect to the infrared spectra (**Figure 5(e)**), it can be observed that the interaction with the quasicrystal and the resulting conformational modifications of dopamine significantly influence the low-frequency region ($< 500$ $cm^{-1}$), where some bands split into more defined and sharper peaks. In contrast, the spectral profile remains mostly unchanged for frequencies below approximately 2000 $cm^{-1}$, indicating that the fundamental skeletal vibrations of the molecule are largely preserved. At higher frequencies, however, new absorption features emerge around 2600, 2800, and 3100 $cm^{-1}$, suggesting subtle alterations in the stretching modes associated with C-H, O-H, and N-H bonds due to adsorption on the quasicrystal surface.

### 3.7 Raman Spectroscopic Investigation of 2D Al QC with Dopamine Functionalization

The interaction between dopamine and 2D Al QC was investigated using Raman spectroscopy. Supporting Information **Figure S8** presents the Raman spectrum associated with 2D Al QC dispersed on the Si wafer. Raman spectrum of pristine 2D Al QC showing major peaks at 298 $cm^{-1}$ (Cu-O Stretching) [40], 438 $cm^{-1}$ (Al-O stretching), 672 $cm^{-1}$ (O-Al-O stretching) [41, 42], 990 $cm^{-1}$, 1359 $cm^{-1}$, 1601 $cm^{-1}$, 1755 $cm^{-1}$, and 2108 $cm^{-1}$.

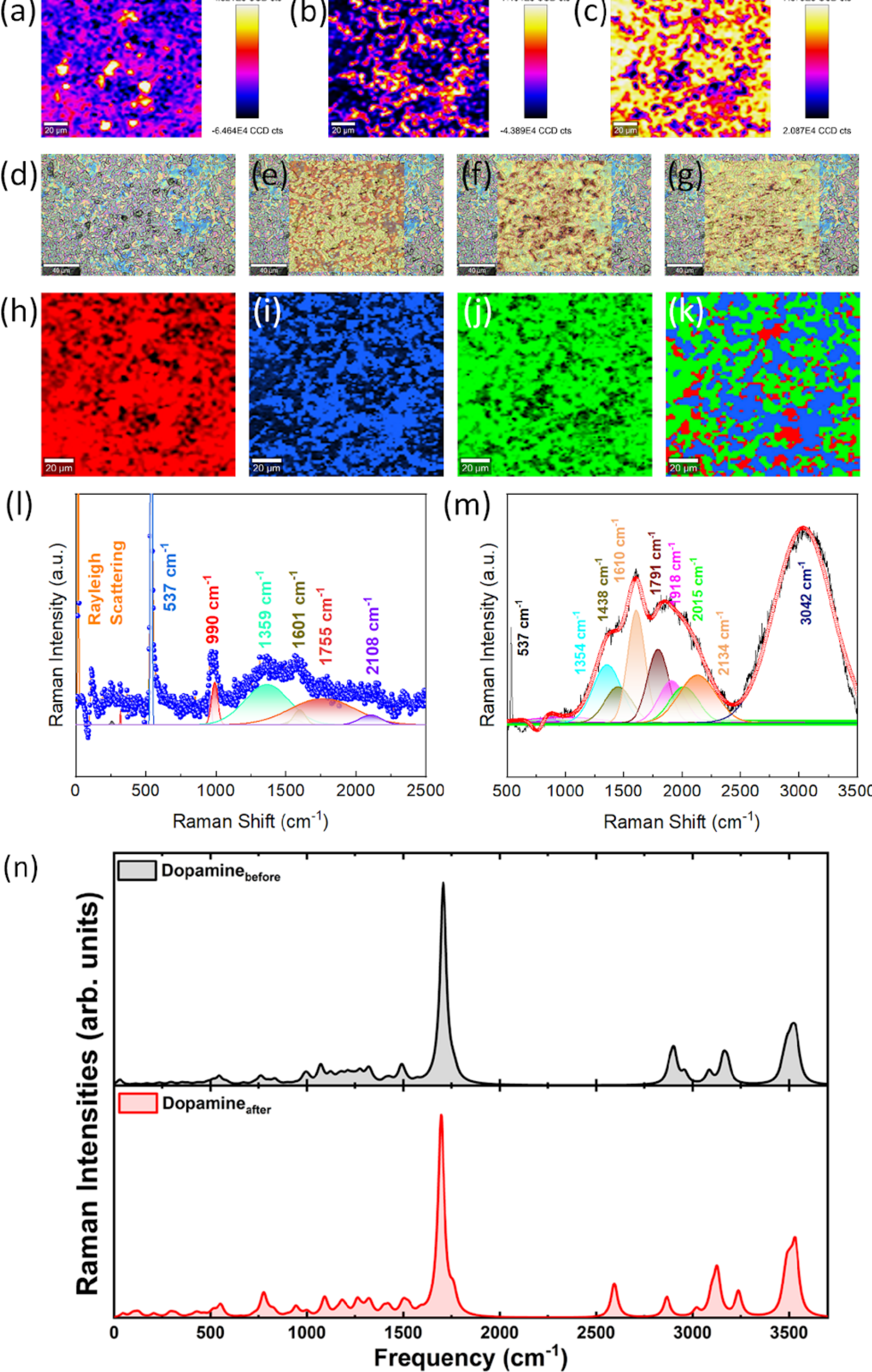

**Figure 6**. (a-b-c) Raman mapping of the 2D Al QC shows Raman signals from the 2D Al QC, Dopamine, and the Si wafer. (d) Optical Image of the 2D Al QC Nanostructure-Dopamine dispersed on Si wafer. (e) Overlay image of **Figure (a)** on **Figure (d)** depicting the localized Raman signal of the 2D Al QC. (f) Overlay image of **Figure (b)** on **Figure (d)** depicting the localized Raman signal of the Dopamine. (g) Overlay image of **Figure (c)** on **Figure (d)** depicting the localized Raman signal of the Si wafer. (h) Raman Spectrum Data corresponding to **Figure (b)** or Dopamine on Si Wafer. (m) Raman Spectrum Data of 2D Al QC-Dopamine on Si Wafer corresponding to **Figure (c)**. (n) Simulated Raman spectra of dopamine before and after adsorption on the quasicrystal (QC) surface. The comparison highlights the vibrational mode shifts and intensity variations associated with structural and electronic changes induced by adsorption.

Raman mapping was performed on a drop-cast sample of 2D Al QC on the Si wafer to investigate the spatial distribution of these vibrational modes. To further explore the possibility of sensing the dopamine molecule, **Figure 6(d)** presents the optical image at 100× magnification of a drop-casted dopamine/2D-Al QC sample on a silicon wafer, accompanied by the accompanying Raman maps in **Figure 6(a)** to **Figure 6(c)**. **Figure 6(e)** to **Figure 6(g)** shows the superimposed Raman signal on **Figure 6(d)** (Optical Image). In these maps, yellow indicates the maximum intensity, brown indicates the minimum, and a mix of both indicates intermediate levels. At 438 $cm^{-1}$, the red patches correspond to 2D-Al QC nanostructures, indicating Al-O bond vibrations associated with Dopamine binding, whereas the silicon substrate appears in purple to blue. The green signal at 537 $cm^{-1}$ in **Figure 6(j)** originates from the silicon substrate, which shows a pronounced Raman peak at this wavelength. At 3042 $cm^{-1}$, dopamine molecules are predominantly associated with 2D-Al QC. **Figure S2c** shows the optical image and Raman map of unaltered 2D-Al QC. The Red (**Figure 5(h)**) and blue images (**Figure 6(i)**) exhibit the signature of robust Raman intensities from 2D Al QC and Dopamine bonds, but the 520 $cm^{-1}$ picture emphasizes the silicon substrate in green. The major involvement of the Al-O bond with the -OH group was observed. Another important aspect investigated was the vibrational response of dopamine, specifically its Raman spectral signatures, to assess the influence of interaction with the quasicrystal surface. **Figure 6(l)** shows the Raman spectrum of the dopamine molecule. **Figure 6(m)** shows the 2D Al QC interaction with dopamine molecules, followed by the spectrum. The following peak shifts are attributed to the interaction with the dopamine molecule and the major involvement of the -OH group.[41] As shown in **Figure 6(n)**, the overall Raman profile preserves the main characteristic peaks of dopamine; however, in the region between 500 and 1500 $cm^{-1}$, several minor bands become more distinguishable and better resolved after adsorption. Moreover, the broad feature with a shoulder observed between 2750 and 3000 $cm^{-1}$ in the isolated molecule appears to split into two distinct peaks, centered approximately at 2600 and 2800 $cm^{-1}$, upon interaction with the QC, indicating subtle modifications in vibrational modes associated with C-H and O-H bond dynamics. In addition, in the spectral region above 3000 $cm^{-1}$, peaks corresponding to the stretching vibrations of the amine (N-H) groups are observed, confirming the molecular integrity of dopamine and highlighting shifts indicative of weak interfacial interactions with the quasicrystal surface.

## 4. Conclusions

In this work, an alloy based on Aluminium quasicrystal has been synthesized. Liquid-phase exfoliation via ultrasonication has converted the bulk alloy into a 2D nanostructure, driven by weak interlayer interactions. The 2D Al QC's optical and structural characteristics have been investigated. Absorbance spectra have confirmed their plasmonic character, and transmission images obtained with a 532 nm CW laser as the light source have also shown this. Raman and FTIR analyses confirmed electrostatic interactions with dopamine molecules. The nature of the 2D Al QC alloy and its optical characteristics after interaction with the dopamine molecule have been investigated using DFTB methods. Due to the presence of active Aluminium sites on the surface of the 2D Al QC, activation of Al-O and O-Al-O bonds shows a characteristic component. Dopamine sensing has been accomplished using absorbance spectra of optically active 2D Al QCs, establishing a detection range of 300-1100 ppb. In contrast to traditional sensing methods such as UV-Vis, Raman, and FTIR Spectroscopy, a novel method for detecting dopamine concentration is proposed. Using far-field diffraction patterns under 532 nm CW laser illumination, SSPM spectroscopy has clearly demonstrated the quantification of interaction time and level with dopamine. Refractive index and time-evolution-based sensitivities have been calculated, which indicate the level of dopamine mixing. The maximum number of rings is 18, 13, 8, 5, and 1 for different levels of dopamine mixing concentration, with an incoming laser beam intensity of 6.28 $W.cm^{-2}$. It is found that the maximum refractive index-based sensitivity and time evolution-based sensitivity of the 2D Al QC@RhB system for sensing the Dopamine molecule using the SSPM spectroscopy method are 3.56 $cm^2W^{-1}mL^{-1}$ and 10 $mL^{-1}$, respectively. Whereas traditional spectroscopy-based dopamine sensing relies on footprint detection. DFTB modeling of 2D Al QC with dopamine has been performed. Therefore, it has been shown both experimentally and conceptually that SSPM spectroscopy, based on plasmonically active 2D Al

QC, has the potential to be a low-cost visual sensor for organic molecules. Despite significant progress, several challenges remain in plasmon-activated biosensing, including reproducible fabrication of plasmonic substrates, stability of nanostructures under biological conditions, and precise control of hotspot formation. Future development is expected to focus on hybrid plasmonic-photonic structures, two-dimensional material-sensor integration, and machine learning based assisted spectral analysis, which can further enhance the detection sensitivity and selectivity. The integration of plasmonic nanomaterials with microfluidic systems and portable optical devices is also expected to accelerate the development of point-of-care biosensors for clinical diagnostics.

## Acknowledgements

C.S.T. acknowledges DAE Young Scientist Research Award (DAEYSRA), SPARC (Scheme for Promotion of Academic and Research Collaboration) funding, and AOARD (Asian Office of Aerospace Research and Development) grant no. FA2386-21-1-4014, and Naval Research Board. Guilherme S. L. Fabris acknowledges the São Paulo Research Foundation (FAPESP) fellowship (process number 2024/03413-9). Raphael B. de Oliveira thanks the National Council for Scientific and Technological Development (CNPq) (process number 200257/2025-0), Douglas S. Galvão acknowledges the Center for Computing in Engineering and Sciences at Unicamp for financial support through the FAPESP/CEPID Grant (process number 2013/08293-7). We thank the Coaraci Supercomputer Center for computer time (process number 2019/17874-0). The authors acknowledge the Sophisticated Analytical and Technical Help Institute (SATHI) at IIT Kharagpur for STEM (JEM-ARM 300 F2) measurements. D. S. G. also acknowledges support from INEO/CNPq and FAPESP grant . 2025/27044-5.